**Novel magnetoresistance features in HgSe single crystal with low electron concentration**

A.T.Lonchakov, S.B.Bobin, V.V.Deryushkin, V.I.Okulov, T.E.Govorkova, and V.N.Neverov

*M.N. Miheev Institute of Metal Physics of Ural Branch of Russian Academy of Sciences, 620108, Yekaterinburg, Russia*

**Abstract**

For the first time, magnetoresistive properties of the single crystal of HgSe with a low electron concentration were studied in wide range of temperature and magnetic field. Some fundamental parameters of spectrum and scattering of electrons were experimentally determined. Two important features of magnetic transport were found – strong transverse magnetoresistance (MR) and negative longitudinal MR, which can indicate the existence of the topological phase of the Weyl semimetal (WSM) in HgSe. Taking this hypothesis into account we suggest a modified band diagram of the mercury selenide at low electron energies. The obtained results are essential for the deeper understanding of both physics of gapless semiconductors and WSMs – promising materials for various applications in electronics, spintronics, computer and laser technologies.

Weyl and Dirac semimetals have recently attracted great attention as materials that possess strong spin-orbit coupling (SOC) and relativistic electron spectrum. Such materials can be considered to be three-dimensional analogues of graphene. For an WSM phase to exist there should be breaking of either inversion symmetry, as in monophosphides and monoarsenides of tantalum and niobium TaP [1,2], TaAs [3], NbP [4], NbAs [5], or time reversal symmetry, as in the ferromagnetic spinel $HgCr_2Se_4$ [6]. The breaking of inversion symmetry also occurs in mercury selenide – gapless semiconductor that crystallizes into a zinc-blende structure and has inverted electron spectrum in the center of the Brillouin zone. The zinc-blende structure consists of two mutually interpenetrating face-centered cubic lattices with a tetrahedral coordination of atoms. HgSe belongs to the space symmetry group $F\bar{4}3m$ with the lattice constant $a = 6.074$Å and coordination number Z = 4. The band order is inverted because of relativistic effects including SOC [7]. Mercury selenide with low electron concentration can become relevant material in the topological condensed matter physics. In this regard, it should be noted that in HgSe under no annealing condition could the electron concentration be reduced below $n_e \sim 10^{16}$cm$^{-3}$ at 4.2K [7]. Smallness of the electron concentration is essential to reveal the Weyl nodes – features of an energy spectrum of topological nature. The Weyl nodes (or magnetic monopoles) are band touching points in the momentum space, which always come in pairs of opposite chirality [8]. Near the Weyl node, the energy spectrum is a spin non-degenerate cone, unlike Dirac semimetals in which the states in the cone are twofold spin degenerate [9].

To the best of our knowledge, no studies of the magnetoresistive properties of single crystals of HgSe with the minimal $n_e$ in a wide range of magnetic fields and temperatures have been carried out so far. Neither there were any attempts to calculate the band structure of HgSe including SOC in the range of small energies of ~ (100 – 200) meV in different directions of the Brillioun zone, as it was done for TaP [10,11], TaAs [11,12], NbP и NbAs [11] in order to reveal the Weyl nodes. Performing *ab initio* band-structure calculations for HgSe can be justified by the paper [13] in which the possibility of existence of up to 12 pairs of Weyl nodes, and Fermi arcs for the zinc-blende structure was theoretically predicted.

In this paper we report the first results of the investigation of the temperature dependence of resistivity and temperature and magnetic field dependences of transverse and longitudinal magnetoresistance (MR) for the nominally pure single crystal of mercury selenide with $n_e = 2.5 \cdot 10^{16}$ cm$^{-3}$ at $T =4.2$K. This value of $n_e$ is smaller than that of HgSe samples in Ref. [14] in which temperature-dependent electrical properties of HgSe crystals were reported. We studied magnetoresistive effects in the temperature range of (1.8 – 300) K and magnetic fields up to 12T. The standard four-probe method was employed to measure magnetoresistance. The sample under investigation was cut from a homogeneous part of the single-crystal ingot grown by the Bridgman method. Then, it was annealed in selenium vapor for ≈ 600 h at 160°C to decrease the electron concentration [14]. The sample had the shape of a rectangular parallelepiped with dimensions (3.5×0.92×0.65) mm$^3$. Prior to the application of the electric contacts, the sample was polished and etched in the 10% solution of bromine in methanol for 1 min [15]. As a material for electric contacts, we used the eutectic of gallium and indium with the addition of silver. The distance between the potential probes was 0.8 mm. In addition, it should be kept in mind that the current contacts fully covered the end surfaces of the sample.

The magnetic-field dependence of the normalized transverse magnetoresistance $MR = [(\rho_{xx}(B) - \rho_0)/\rho_0] \times 100\%$ is shown in Fig. 1 (here $\rho_{xx}(B)$ is MR at $B \perp I$, $\rho_0$ is the resistivity in zero magnetic field, $I$ is the current). It is seen that at all temperatures the sample resistivity increases with magnetic field without any sign of saturation. The maximum value of $MR$ is ≈ 7200% at 40K in a field of 12T. The value of the MR is 200% at room temperature in 9T. It was determined that at temperatures on Fig. 1 the MR has a parabolic behavior except at 40K where it changes from parabolic to a linear dependence above 7T.

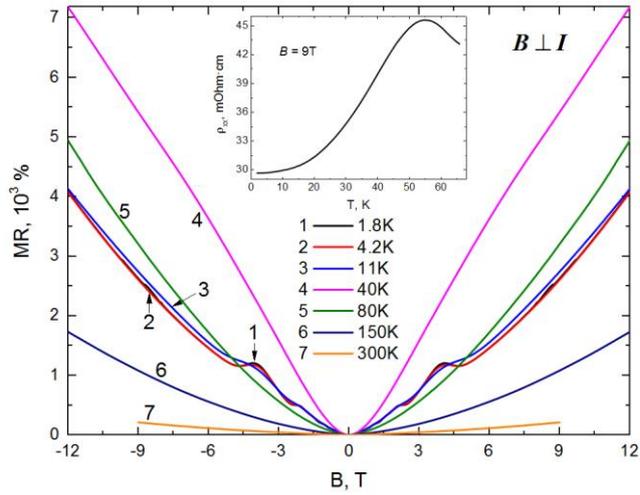

*Fig.1. Magnetic field dependence of the normalized transverse MR (B ⊥ I) measured at different temperatures for the HgSe single crystal. Inset shows the temperature dependence of the transverse MR at 9T, which demonstrates the maximum at 55K.*

The Hall mobility $\mu_H$ of the sample at 4.2K is $1.5 \cdot 10^5$ cm$^2$/V·s. Despite the fact that it decreases to $2 \cdot 10^4$ cm$^2$/V·s at 300K, the MR continues to increase at $\omega_c \tau_{tr} \gg 1$, where $\omega_c = eB/m_c$ is the cyclotron frequency, $\tau_{tr} = \mu_H m_c/e$ is the transport scattering time, $m_c$ is the effective mass of electron, $e$ is the electron charge. Discovery of a large positive magnetoresistance in HgSe with low electron concentration in a wide temperature range is an important result of this report. Materials that possess large positive MR can find practical application in magnetic sensors and magnetic memory devices. A brief overview of the reports on various aspects of positive MR is presented in [2, 16]. The problem of positive MR comprises giant MR, colossal MR, positive linear MR in inhomogeneous systems, and quantum linear MR. For systems with a high carrier mobility, a large magnetoresistance effect is typical of topological insulators [17], Dirac semimetals [18, 19] and WSMs [1, 2, 4, 5]. Generally, according to [18, 19], in topological materials with high mobility, $\tau_{tr}$ is much larger than the quantum scattering time $\tau_q$. This can be caused by a certain mechanism of protection from the large angle scattering in the absence of magnetic field. Application of magnetic field results in lifting the protection from backscattering and the MR greatly increases. We have determined $\tau_{tr}$ = $2.7 \cdot 10^{-12}$ s and $\tau_q = \hbar/2\pi k_B T_D$ =$3.7 \cdot 10^{-13}$ s (here $\hbar$ is the Plank constant, $k_B$ is the Boltzmann constant) using the values of $m_c$ and Dingle temperature $T_D$ from the experiment, as will be demonstrated below. Since $\tau_{tr} \gg \tau_q$, one can conclude that in our sample small angle scattering is predominant, *i.e.*, according to [18], the condition for the appearance of large magnetoresistance effect is satisfied. As mentioned above, we found that the positive MR in HgSe is a nonmonotonous function of temperature. As seen in the inset of Fig. 1, the curve $\rho_{xx}(T)$ measured in the magnetic field of 9T shows a maximum at 55K and saturates at low temperatures. A similar behavior of MR was observed in WSM TaP [2], TaAs [20,21], and NbAs [22], yet, it has not received a satisfying theoretical explanation.

Shubnikov-de Haas (SdH) oscillations of the transverse magnetoresistance $\Delta\rho_{xx}(B)$ with the frequency $F = 2.6T$ were obtained by subtracting smooth background from $\rho_{xx}(B)$ at $T = 1.8$, 4.2, and 11K. In this brief report, we analyze only the amplitude of SdH oscillations, which according to the Lifshitz-Kosevich (LK) theory for 3D systems is given by formula [23]:

$$A(B,T) = \frac{5}{2}\left(\frac{B}{2F}\right)^{\frac{1}{2}} \frac{\lambda}{\sinh\lambda} \exp(-\lambda_D), \tag{1}$$

where $\lambda = 2\pi^2 k_B T/\hbar\omega_C$, $\lambda_D = 2\pi^2 k_B T_D/\hbar\omega_C$. Using (1), one can determine $m_c$ and $T_D$ from the temperature and magnetic field dependences of the oscillations amplitude, respectively. The cyclotron effective mass of electron was obtained from the analysis of the SdH oscillations in the following way. First, we plotted envelope curves $f_{max}(B^{-1})$ and $f_{min}(B^{-1})$ on maximums and minimums of $\Delta\rho_{xx}(B^{-1})$, respectively (Fig. 2a). Then we determined the amplitude of oscillations as $A(B_{max}^{-1},T) = |\Delta\rho_{xx}(B_{max}^{-1}) - f_{\min}(B_{max}^{-1})|$ and $A(B_{min}^{-1},T) = |\Delta\rho_{xx}(B_{min}^{-1}) - f_{\max}(B_{min}^{-1})|$. In Fig. 2b, the amplitude is marked with symbols, and their exponential fitting functions are plotted as solid lines. Next, we obtained the ratio of the fitting functions for $T_1=1.8K$ and $T_2=4.2K$ (dashed line in the insert of the Fig. 2b). It was fitted with theoretical curve using expression:

$$\frac{A(B,T_1)}{A(B,T_2)} = \frac{T_1}{T_2}\frac{\sinh\lambda(B,T_2)}{\sinh\lambda(B,T_1)}, \tag{2}$$

where $m_c$ is the fitting parameter (solid curve in the inset of Fig. 2b). The discrepancy between the experiment and LK theory for this curve did not exceed 4%. The theoretical curve corresponds to the effective mass $m_c = 0.0329 m_0$, where $m_0$ is the mass of a free electron. It should be noted that the effective electron mass in HgSe with $n_e \sim 10^{16} cm^{-3}$ was obtained from the experiment for the first time. To determine $T_D$, we used the magnetic-field dependence of the oscillation amplitudes $A(B,T)$ at $T = 1.8$ and 4.2K (solid lines in Fig. 2b). From the expression (1) one can obtain $T_D = -(\hbar e/2\pi^2 k_B m_c)tg\varphi$, where $\varphi$ is the slope in the semilog plot $\left[A(B,T)B^{\frac{1}{2}}\sinh\lambda(B,T)\right]$ versus $B^{-1}$. The experimental curves obtained from Fig. 2b are shown in Fig. 2c as solid lines. Dashed lines are their linear fitting, which gives $T_D= 3.39$ and 3.29K for $T = 1.8$ and 4.2K, respectively. Also, from the SdH oscillations we obtain a Fermi surface area of $A_F = 2\pi eF/\hbar = 2.48\cdot 10^{-4}$ Å$^{-2}$, the Fermi wave vector $k_F$ is $(A_F/\pi)^{\frac{1}{2}} = 8.9\cdot 10^{-3}$ Å$^{-1}$, and the Fermi velocity $v_F$ is $(\hbar k_F/m_c) = 3.3\cdot 10^5$ m/c.

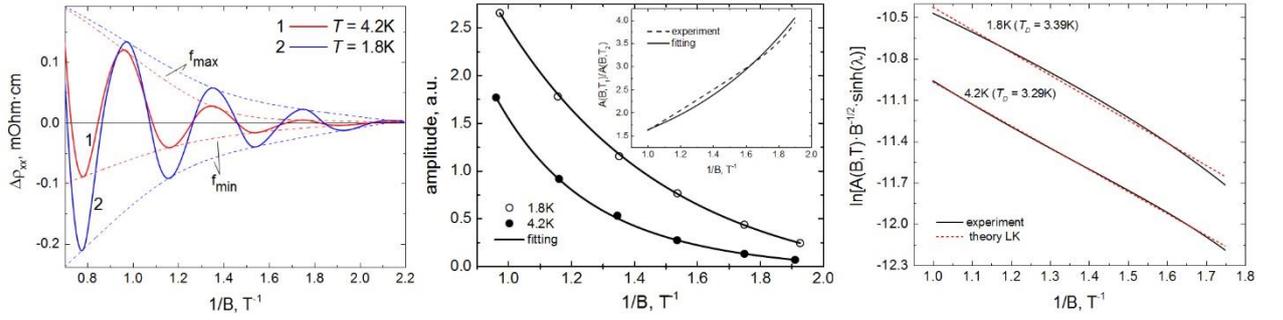

*Fig.2. SdH oscillations of the transverse MR and their analysis. (a) The oscillatory component $\Delta\rho_{xx}$ of MR for T =1.8 and 4.2K. Dashed lines are upper and lower envelopes $f_{max}$ and $f_{min}$ for $\Delta\rho_{xx}$. (b) Field dependence of the oscillations amplitude (symbols) and its exponential fitting (solid lines). Inset shows the ratio of the fitting functions for $T_1$=1.8K and $T_2$= 4.2K versus reciprocal magnetic field (dashed line). The solid line in the inset is its theoretical fitting by LK formula. (c) Dingle plots of $ln\left[A(B,T)B^{\frac{1}{2}}sinh\lambda(B,T)\right]$ versus $B^{-1}$, used to determine the Dingle temperature $T_D$ and quantum scattering time $\tau_q$.*

The magnetic field dependence of the normalized longitudinal magnetoresistance (LMR) ($B \parallel I$) for the HgSe sample is shown in Fig. 3. It is seen that upon experimental conditions, LMR increases with *B*, reaches maximum, and then decreases demonstrating the effect of negative MR. The maximum value of positive MR is 70%. The negative MR also reaches the maximum value of 70% at 20K and 10T. At low temperatures, the SdH oscillations of LMR are present. Therefore, the appearance of minimum of the negative LMR curve at temperatures 2, 4.2, 10 and 20 K can be associated with the LMR increase in the ultra-quantum limit.

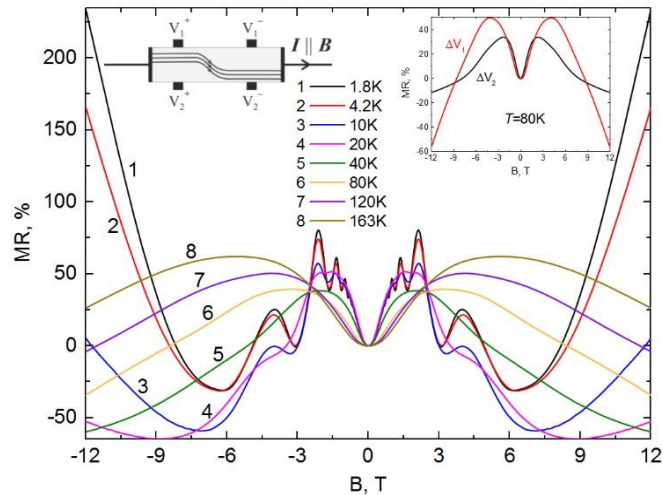

*Fig.3. Magnetic field dependence of the normalized LMR ($B \parallel I$) measured at different temperatures in the HgSe single crystal. Left inset shows the sketch of the sample with*

*current density distorted by parallel magnetic field [26] and with two pairs of potential probes located symmetrically relative to the sample center. Right inset shows the field dependence of LMR measured at 80K separately on this probes with the potential differences $\Delta V_1$ and $\Delta V_2$.*

When studying the MR for *B ∥ I*, other sources of the negative LMR should be considered because they can affect the magnetic-field dependence of LMR [3]. As was noted above, the current contacts fully covered end surfaces of the sample. This precaution [3] effectively prevents the results of measurements from being distorted by geometrical or size effects like current jetting [24, 25]. The negative MR can also originate from inhomogeneity of current density along the sample in a strong magnetic field [26], as shown in the left inset of Fig. 3. As follows from Ref. [26], such distortion effect can be significantly reduced by averaging the voltage values measured by potential probes localized strictly against each other on opposite lateral surfaces of the sample $\overline{\Delta V} = [(V_1^+ - V_1^-) + (V_2^+ - V_2^-)]/2$, as shown in the right inset of Fig. 3. The thus averaged curves of LMR are shown in Fig. 3. An example of non-averaged LMR curves, measured separately at different pairs of potential probes, is shown in the right inset of Fig. 3.

The LMR detected in HgSe (Fig.3) is both qualitatively and quantitatively very much alike the LMR in the WSMs such as TaP [2], TaAs [3, 20], NbP [4], and NbAs [22]. In these reports, the increase of LMR in low magnetic field is explained by a weak antilocalization (WAL), whereas the negative LMR results from a chiral anomaly. It means that in parallel electric and magnetic fields, the number of Weyl fermions of positive and negative chirality are not separately conserved [27]. Thus, pumping of electrons between the Weyl nodes with the opposite chirality takes place. This results in the positive contribution to the conductivity that has a quadratic dependence on magnetic field [28, 29]. We should note that the negative LMR in the WSMs, being a consequence of the chiral anomaly, is considered in the literature as a fingerprint of a Weyl semimetal phase.

Thus, we can claim the discovery of a negative LMR in the single crystal of HgSe that resembles the negative MR induced by the chiral anomaly in the WSMs. In combination with the peculiarities of the positive transverse MR, it gives reason to assume the existence of WSM phase in HgSe, i.e., the existence of at least two pairs of the Weyl nodes (labeled as W1) with the opposite chirality near the Fermi level. In this case, simplified band diagram for our sample at low energies could be like one in Fig. 4. Here, along with the well-known gapless spectrum of HgSe [7] in the center of the Brillioun zone, a W1-type Weyl cone with the node locations ($\varepsilon_{W1}, k_{W1}$) is supposed. The Weyl cone with the opposite chirality is not shown to simplify the scheme. Note that, according to the *ab initio* calculations, a band spectrum of the kind is realized in the WSM NbP [30]. The only difference is that in NbP the trivial bulk spectrum is that of a classical semimetal rather than a gapless semiconductor, as it is the case for HgSe (Fig. 4). Let us note that in both NbP and HgSe, the energy $\varepsilon_{W1}$ is negative with respect to the Fermi energy. As we suppose, the case in Fig. 4 corresponds to our sample, in which rather a long annealing could lead to the depopulation of conductivity band $\Gamma_8$. Therefore, the kinetic properties of the sample are determined by the electron pocket at the Fermi level, which is a Weyl cone. For the Dirac spectrum, the Fermi energy $\varepsilon_F$ taken relative to the node energy $\varepsilon_{W1}$ is $m_c v_F^2 = 18.3$meV. As it is seen from Fig. 4, the Fermi level of this system is close to that of intrinsic HgSe. In this case, the concentration of electrons $n_e \cong 2.5 \cdot 10^{16}$см$^{-3}$ corresponding to $\varepsilon_F$ can be considered as extremely low in HgSe.

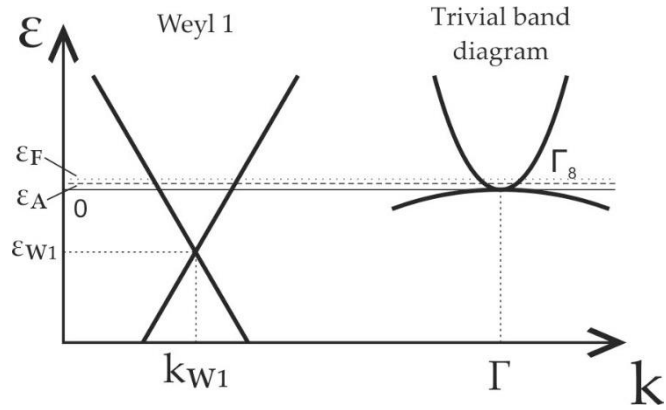

*Fig.4. The scheme of the band structure of HgSe at low electron energy, which represents the hypothesis of the existence of Weyl nodes of W1- type, with energy $\varepsilon_{W1}$ near the Fermi level. To simplify the scheme, the Weyl node with the opposite chirality is not shown. Dash line marks the energy of acceptor level $\varepsilon_A \cong \varepsilon_F$ (dotted line), near the top of the valence band.*

The coexistence of the topological electron pocket and trivial gapless spectrum of the mercury selenide enables us to suggest the interpretation of the unusual temperature dependence $\rho_0(T)$ that shows two maximums (Fig. 5): "low-temperature" at $T_1 \approx 5$K and "high-temperature" at $T_2 \approx 45$K. Let us discuss the possible nature of such unusual behavior of $\rho_0(T)$ within the classical physics of gapless semiconductors [7]. It is known that in an intrinsic gapless semiconductor, the electron concentration is a power function of temperature: $n_e \sim m_e m_h^{\frac{1}{2}} T^{\frac{3}{2}}$, where $m_e$ and $m_h$ are the effective mass of electron and hole, respectively.

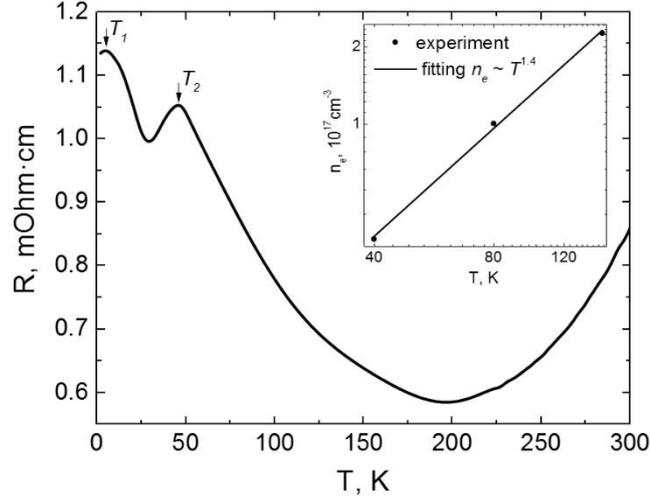

*Fig. 5. Temperature dependence of resistivity in the absence of magnetic field. The arrows indicate the positions of maximums of $\rho_0(T)$ at $T_1=5K$ and $T_2=45K$. Inset shows temperature dependence of electron concentration (symbols) and its fitting with the power function $T^{1.4}$.*

From the Hall measurements, we obtained $n_e \sim T^{1.4}$ below 150K where $\rho_0(T)$ increases (inset of Fig. 5), which is close to the predicted power law dependence. A small difference of exponents indicates to the contribution of the impurity conductivity owing to intrinsic donor-type defects, like for example, vacancies of Se. Thus, the maximums of the $\rho_0(T)$ appear on the background of the temperature change of $n_e$. In this case, their existence can be explained as the effect of acceptors, which, similarly to the donor-type defects, are always present in HgSe at small concentration [7, 14]. If the acceptors are few and the acceptor level is smeared a little, electrons will be captured by acceptors and become practically bound. Therefore, with an increase in temperature, the electron concentration should reduce over some temperature interval. As the temperature is raised further, electrons begin to be excited from the valence band to the conduction band and, hence, the electron concentration rises. As a result, a minimum of $n_e(T)$ must be observed. The position of the minimum of $n_e(T)$, and, consequently, maximum of $\rho_0(T)$, depends on the energy of the acceptors $\varepsilon_A$ and is determined from the condition $k_B T_{min} \approx 0.25 \varepsilon_A$, if $\varepsilon_F \cong \varepsilon_A$ [7]. In the effective mass method, the acceptor level is always attached to the top of the valence band. On the other hand, for the Weyl cone, the energy $\varepsilon_A$ should be determined relative to the node energy $\varepsilon_{W1}$ (Fig. 4). The effect of the nonmonotonic dependence of the electron concentration on temperature will take place for both W1- type Weyl cone and trivial gapless phase of HgSe. From the condition of the $\rho_0(T)$ maximum, we obtain $\varepsilon_A \approx 2$ meV for $T_1$ and $\varepsilon_A \approx 16$ meV for $T_2$. The former energy value of the acceptor state correlates well with the value for the gapless semiconductors [7], whereas the latter value, as expected (Fig.4), turns out close to the energy $\varepsilon_F$ taken relative to the Weyl node energy. Thus, we can assume that the appearance of the high-temperature maximum in $\rho_0(T)$ is connected to peculiar topological properties of HgSe.

In Conclusion, for the first time, the magnetoresistive properties of the HgSe single crystal with the low concentration and high mobility of electrons were studied in a wide range of magnetic field and temperature. We discovered a large positive transverse magnetoresistive effect reaching up to 7200% with no sign of saturation, which is valuable for practical applications. We established that the field and temperature dependences of the transverse MR have features typical of WSMs, which are considered to be promising materials for novel computer technologies. From the analysis of the temperature and magnetic field dependences of the amplitude of SdH oscillations of the transverse MR, the effective mass and the Dingle temperature were determined for HgSe with $n_e \sim 10^{16}$cm$^{-3}$ for the first time. For $B \parallel I$, a negative magnetoresistance was detected which can be related to a chiral anomaly arising in WSMs. It is established that the unusual magnetoresistive properties of the HgSe single crystal are accompanied by the anomalous temperature dependence of resistivity. To interpret the observed features of the MR and electrical resistivity, we have suggested a modified band model for the mercury selenide at low electron energies. In this model, along with the trivial gapless spectrum at the Γ point of the Brillouin zone, there exists a phase of the Weyl semimetal with the Weyl node located by 18 meV lower than the Fermi level in the intrinsic mercury selenide. We consider this brief report as a starting point in studying HgSe as potential candidate in the family of WSMs.

**Acknowledgments.** The work was performed as a part of the state task FASO Russia (subject "Electron", No. 01201463326) with partial support by the RFBR (project No. 16–32-00131), Fundamental Research Program of UB RAS 2015–2017 (project No 15–17-2–32).